# Valley-Selective Phonon-Magnon Scattering in Magnetoelastic Superlattices

Liyang Liao[1], Jorge Puebla[2,*], Kei Yamamoto[3,2], Junyeon Kim[2], Sadamichi Meakawa[2,3,5], Yunyoung Hwang[1,2], You Ba[2], and Yoshichika Otani[1,2,4*]

[1]*Institute for Solid State Physics, University of Tokyo, Kashiwa 277-8581, Japan*
[2]*Center for Emergent Matter Science, RIKEN, Wako, Saitama 351-0198, Japan*
[3]*Advanced Science Research Center, Japan Atomic Energy Agency, Tokai 319-1195, Japan*
[4]*Trans-scale Quantum Science Institute, University of Tokyo, Tokyo, Japan*
[5]*Kavli Institute for Theoretical Sciences, University of Chinese Academy of Sciences, Beijing 100190, China*

*e-mail: jorgeluis.pueblanunez@riken.jp, yotani@issp.u-tokyo.ac.jp

**Abstract:**

**Phonons and magnons are engineered by periodic potential landscapes in phononic and magnonic crystals, and their combined studies may enable valley phonon transport tunable by the magnetic field. Through nonreciprocal surface acoustic wave transmission, we demonstrate valley-selective phonon-magnon scattering in magnetoelastic superlattices. The lattice symmetry and the out-of-plane magnetization component control the sign of nonreciprocity. The phonons in the valleys play a crucial role in generating nonreciprocal transmission by inducing circularly polarized strains that couple with the magnons. The transmission spectra show a nonreciprocity peak near a transmission gap, matching the phononic band structure. Our results open the way for manipulating valley phonon transport through periodically varying magnon-phonon coupling.**

Periodic potential landscapes can modify the coupling between two quasiparticles, leading to various emergent phenomena. For example, Moiré superlattices can support superconducting [1,2] or quantum anomalous Hall states [3,4] by coupling two layers and forming entangled spin and valley. Bosonic excitations, such as phonons and magnons, are important quasiparticles in solids and have inherent coupling [5–8]. The band structure and wave function of phonons and magnons can be tailored by periodic landscapes in phononic [9–12] and magnonic crystals [13–16]. In particular, similar to the electronic valleys in graphene and transition metal dichalcogenide (TMD) [17,18], honeycomb phononic crystals also have the valley degrees of freedom [19,20]. Honeycomb phononic crystals with broken inversion symmetry can mimic the TMDs, bringing about the phononic valley Hall effect [20] and phononic topological edge states [19,21–23]. Based on magnon-phonon coupling, theoretically, magnetic-field-tunable topological chiral phonons have been proposed [24–26]. However, experimental demonstration of valley phonon transport manipulated by magnon-phonon coupling has yet to be achieved.

Phonons at different valleys are time-reversal pairs. Since magnetism breaks the time-reversal symmetry, magnons may interact differently with the phonons in the two valleys. As shown in Fig. 1(a), for a honeycomb phononic crystal with sites A and B in the unit cell, there is a pair of valleys K, K', in the Brillouin zone. When the spatial inversion symmetry is broken by inequivalent A and B sites, and an out-of-plane magnetic field component also breaks the time-reversal symmetry, a valley-selective magnon-phonon interaction can emerge, as presented in Figs. 1(b) and (c); under an upward magnetic field, the phonon in the K valley interacts with the magnons, while the K' valley does not. The situation reverses when the magnetic field is in the opposite direction, and the interaction exists only in the K' valley.

In this Letter, we report the valley-selective phonon-magnon scattering in magnetoelastic superlattices [Fig. 1(d)]. For forward and backward surface acoustic waves (SAW), phonons fall into different valleys and are scattered by the magnetization with different amplitudes, leading to different SAW transmissions, namely nonreciprocal transmission [27,28]. We proposed a circular polarization of the strain field (hereafter strain polarization), which couples with the magnons through magnetoelastic coupling. The strain polarization possesses opposite chirality for the two valleys, resulting in different magnon-phonon coupling strengths, allowing the valley-selective phonon-magnon scattering. Our results demonstrate the manipulation of valley phonon transport via magnon-phonon coupling and provide a new avenue for the coupling between out-of-plane magnetization and SAW [29,30].

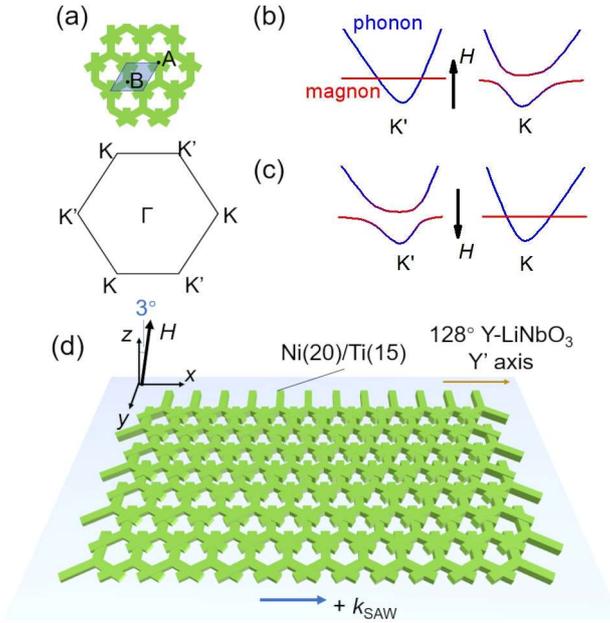

FIG. 1. (a) Unit cell of the honeycomb superlattice and the corresponding Brillouin zone. (b-c) Schematic of the phonon bands and the magnon bands at the K and K' valleys in a symmetry-breaking honeycomb superlattice under an upward magnetic field (b) and a downward magnetic field (c). (d) Honeycomb superlattice made of Ni(20)/Ti(15) in the measurement setup. Magnetic field $H$ is tilted 3° from the out-of-plane $z$ direction in the $xz$-plane, and SAW is excited along the Y'-axis of the 128° YX LiNbO$_3$ substrate, defined as $x$ direction.

We fabricated mirror-symmetry-breaking honeycomb-shaped superlattices using Ni (20 nm)/Ti (15 nm) bilayer film on a 128° Y-cut LiNbO$_3$ substrate, as shown in Fig. 1(d). The edge length is 770 nm, and the SAW wavelength corresponding to the K or K' point is 2 μm. A detail of the superlattice geometry is shown in Supplemental Material (SM) Sec. 1. The SAW, excited by interdigital transducers (IDT), travels along the substrate Y'-axis for more efficient hybridization at the valleys (SM Sec. 2) [31].

While scanning a magnetic field $H$, the transmission parameters S12/S21 are measured by a vector network analyzer (VNA) with a time-domain gating technique. The $z$-axis is normal to the substrate, and the SAW propagation direction is the $x$-axis. If not specified, the applied magnetic field is 3° tilted from the out-of-plane $z$-axis in the $xz$-plane. Superlattice orientation dependence are studied by patterning rotated superlattices while fixing the SAW propagation direction on the substrate.

Figure 2 shows the transmission measurement results. To compare results in different devices, we normalize the transmission in each device using the transmission at $\mu_0 H = 0.6$ T, where no magnetization-related SAW absorption exists. In the K-oriented device (KD), $+k$ SAW propagates along the K direction [Fig. 2(a)], whose transmission is represented by S12 (red), and $-k$ SAW along the K' direction, represented by S21 (black). The magnetic hysteretic responses are obtained by scanning the external field between 0.6 T and $-0.6$ T. An absorption peak appears at ~ 0.3 T, as indicated by the blue arrow. The transmission signal S12 is larger than S21, showing a nonreciprocity between the $+k$ and $-k$ SAW propagations. At the negative field side, the absorption peak appears at ~ $-0.3$ T, with S12 < S21, showing a nonreciprocity with an opposite sign. These absorption peaks and nonreciprocity show no difference in positive-to-negative and negative-to-positive field scans. Absorption peaks at low fields below 0.3 T show butterfly-shaped hysteresis and no nonreciprocity. This may be related to the magnetic resonance under the superlattice-induced anisotropy field in the sample (SM Sec. 3) [31]. We hereafter focus on the high-field peaks showing nonreciprocity. When the superlattice is rotated by 180° while keeping the $+k$ SAW propagating direction, the $+k$ SAW is along the K' direction, and the $-k$ SAW in the K direction. We name it the K'-oriented device (K'D) [Fig. 2(b)]. Nonreciprocal absorption peaks appear in the same field but with S12 < S21 at positive and S12 > S21 at negative fields. The nonreciprocity signal takes an opposite sign in the K'D compared with the KD, revealing that the superlattice geometry governs the nonreciprocity.

As a control experiment, we rotated the superlattice by 90° so the SAW propagates along the M direction (M-oriented device, MD). In this configuration, the lattice preserves the mirror symmetry about the $xz$-plane. The S12 and S21 are nearly identical, showing no noticeable nonreciprocity in this configuration. The normalized nonreciprocal absorption $\Delta P_n = (\Delta P_+ - \Delta P_-)/P$, where $\Delta P_\pm$ is the SAW absorption power for $\pm k$ and $P$ is the transmission power at 0.6 T, is plotted in Fig. 2(d). Under the negative field, the nonreciprocal absorption is negative in KD (red), positive in K'D (blue), and close to zero in MD (green). The $-k$ phonons in KD and the $+k$ phonons in K'D are both in the K' valley, and both have stronger phonon scattering, represented by the more significant SAW attenuation. The K valley phonons, corresponding to the $+k$ phonons in KD and the $-k$ phonons in K'D, have a weaker phonon scattering. Nonreciprocity is absent in the M direction, which respects the mirror plane symmetry. When the magnetic field is turned opposite, the nonreciprocal absorption reverses its sign for both KD and K'D, showing that the $+k$ phonons in KD and $-k$ phonons in K'D attenuate more significantly due to a more pronounced phonon scattering in the K valley than the K' valley. The observed nonreciprocity thus originates from a valley-selective phonon-magnon scattering, where the magnetic field direction controls the selectivity over the K and K' valleys.

To further clarify the role of the honeycomb lattice in the observed nonreciprocity, we design a zigzag-like lattice with mirror symmetry breaking [Fig. 2(e)]. The L and R lattices have broken mirror symmetry about the $xz$-plane, which *allows* the zigzag-like lattice to have nonreciprocity. However, as shown in Fig. 2(f), the nonreciprocal absorption in both R (red) and L (blue) lattices is more than four times smaller than the honeycomb lattice. One underlying reason is that the zigzag-like lattice has no valley degree of freedom since its Brillouin zone corners are all equivalent. ± $k$ SAW will excite the same phonon states at M in the square Brillouin zone. Hence, the broken $xz$ mirror symmetry is insufficient to generate large nonreciprocity, and the sizable nonreciprocity observed in the honeycomb lattice benefits from the valley degrees of freedom.

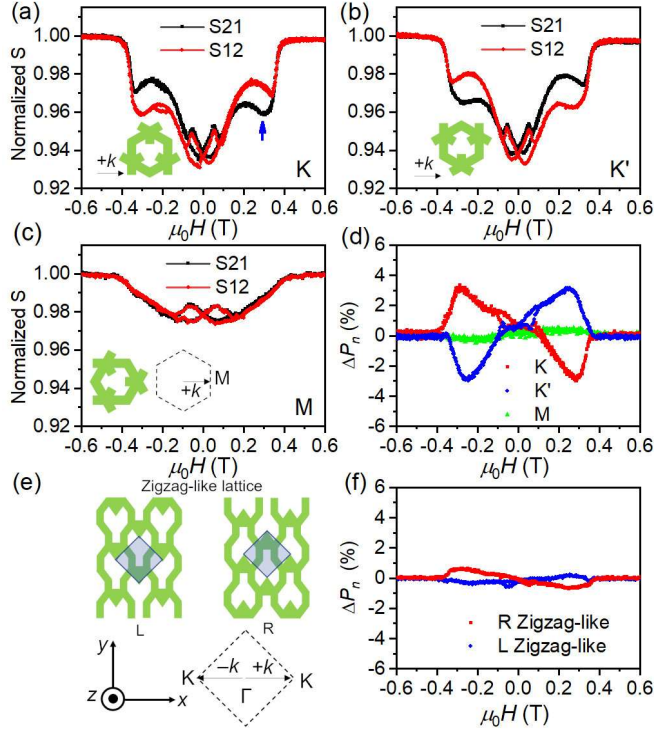

FIG. 2. (a-c) Normalized transmission parameters S12 (red) and S21 (black) as functions of magnetic field $\mu_0H$ for the honeycomb lattice in K-orientated (a), K'-orientated (b), and M-orientated (c) devices. (d) Nonreciprocal absorption $\Delta P_n = (\Delta P_+ - \Delta P_-)/P$ as a function of magnetic field $\mu_0H$ along K (red), K' (blue) and M (green) directions. (e) The geometry of the square lattice with lateral inversion symmetry breaking. (f) Nonreciprocal absorption $\Delta P_n$ for R (red) and L (blue) Zigzag-like lattices with opposite symmetry.

To understand the role of magnetic structure on the valley-selective phonon-magnon scattering, we study the out-of-plane magnetic field angle dependence of the transmission in KD. The measurement geometry is shown in Fig. 3(a), where $\theta_H$ is the magnetic field tilting angle and $\theta$ is the angle between the magnetization and the $z$-axis. Due to the easy-plane anisotropy of the Ni film, under moderate magnetic field, $\theta$ can be significantly different from $\theta_H$. We focus on the $\theta_H$ dependence in the $xz$-plane to avoid additional nonreciprocity from the conventional mechanisms, generated by the $y$-component of magnetization [27,32,33]. After scanning the magnetic field at each $\theta_H$, the peak value of the nonreciprocal absorption $\Delta P_n$ as a function of $\theta_H$ is plotted in Fig. 3(b). The nonreciprocal absorption is almost zero at $\theta_H = 0°$ and nearly symmetric about $\theta_H = 0°$. When the field is tilted away from $\theta_H = 0°$, the nonreciprocal absorption initially increases, peaks at approximately ±3°, and then decreases gradually.

The predominant magnetic field-dependent SAW absorption mechanism in unpatterned Ni film is magnetoelastic coupling due to longitudinal strain $\varepsilon_{xx}$ [33,34].

The magnon-phonon coupling due to the longitudinal strain takes a maximum when the magnetization is directed at 45° from the SAW wavevector, and becomes zero when the magnetization is parallel or perpendicular to the wavevector [34,35]. Therefore, with $\varepsilon_{xx}$ only, the magnon-phonon coupling would be minimized at $\theta = 0°$ and peaked when $\theta = 45°$ at the resonance condition. Meanwhile, the valley phonons can excite rotating displacements in the xy-plane [22], leading to an additional shear strain component $\varepsilon_{xy}$, whose phase is shifted from $\varepsilon_{xx}$. Assuming the dominant terms in magnetoelastic coupling come from these two strain components, one can get

$$\Delta P_n \propto b^2 L \sin 2\theta \sin\theta ,  \quad (1)$$

where $b$ is the magnetoelastic coupling constant, and

$$L = \langle \varepsilon_{xx}(d\varepsilon_{xy}/dt) - \varepsilon_{xy}(d\varepsilon_{xx}/dt) \rangle ,  \quad (2)$$

with the angled bracket representing time-averaging. The detailed derivation is given in SM Sec. 4 [31]. $L$ represents a circular polarization of the strain field. It is odd in frequency $\omega$, while any physical observable of classical wave should be even in $\omega$, $k$, so that it should also be odd in $k$, contributing to the nonreciprocity. Sign change in $\theta$ results in the same $\sin 2\theta \sin\theta$, so that the nonreciprocity given by Eq. (1) is symmetric about $\theta_H = 0°$. Equation (1) gives a maximum $\Delta P_n$ at $\theta = \pm 54.7°$, slightly off $\pm 45°$.

To find the magnetization direction $\theta$ at the resonance condition under a given $\theta_H$, we measured the saturation magnetic field in our superlattice using the anomalous Hall effect, and obtained saturation magnetization $M_S \sim 2.8 \times 10^5$ A/m (Fig. S5) [31]. Based on a marcospin model, we calculate the magnetization direction $\theta$ at which the ferromagnetic resonance (FMR) frequency $f_{FMR}$ meets the SAW frequency $f_{SAW}$; the detailed discussion is given in SM Sec. 5 [31]. Then $\sin 2\theta \sin\theta$ is plotted as a function of $\theta_H$ in Fig. 3(c). The curve reproduces the key features of the experimental results in Fig. 3(b), including the dip at $\theta_H = 0°$, the peak at $\sim 3°$, and the gradual decrease when $\theta_H$ becomes larger. The dip at $\theta_H = 0°$ is due to the fully out-of-plane magnetization $M$, the peaks at $\theta_H \sim \pm 3°$ correspond to $\theta = \pm 54.7°$, and when $|\theta_H| > 3°$, $|\theta|$ gets greater than 54.7°, causing the gradual decrease in $\sin 2\theta \sin\theta$, as illustrated in Fig. 3(c). The good agreement between the calculated curve and the experimental results supports the magnetoelastic coupling as the primary magnon-phonon coupling mechanism in the valley-selective phonon-magnon scattering.

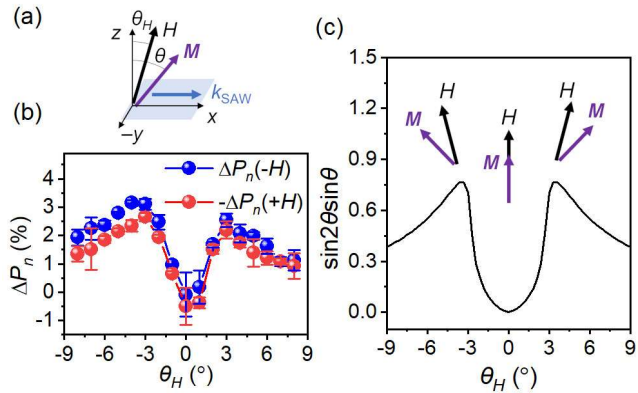

FIG. 3. (a) Schematic of the magnetization angle $\theta$ under the external field $H$ and the easy-plane anisotropy. (b) Peak absolute value of $\Delta P_n$ as a function of the magnetic field tilting angle $\theta_H$. The blue and red data points correspond to negative and positive fields, respectively. (c) Calculated $\sin 2\theta \sin\theta$ at resonance condition $f_{FMR} = f_{SAW}$ as a function of $\theta_H$. Illustrations show the direction of magnetization $M$ at the resonance condition for $\theta_H = 0°, \pm 3°$.

We finally study the nonreciprocal transmission spectra and analyze their relationship with the phonon bands. We focus on the peak condition for $\Delta P_n$ at $\pm 0.28$ T with $\theta_H = 3°$, allowing us to use the ratio $\eta = (\Delta P_+ - \Delta P_-)/(\Delta P_+ + \Delta P_-)$ as a sensible measure of nonreciprocity. Figure 4(a) displays $\eta$ in KD at $\pm 0.28$ T (blue and red points), together with the normalized transmission power $P/P_{w/o}$ at 0.6 T (green curve), as a function of SAW frequency $f$. Here, $P$ and $P_{w/o}$ represent the transmission power with and without superlattice, respectively, detailed in SM Sec. 1 [31]. $P/P_{w/o}$ has a dip at 1.815 GHz, characterizing the gap in the phonon frequency spectrum induced by the lattice. The evolution of $\eta$ is closely related to this spectral gap. When $f$ is far from the gap frequency of 1.815 GHz, $\eta$ is small, but it changes dramatically across the gap. At $\mu_0 H = -0.28$ T, $\eta$ exhibits a negative value at $f = 1.80$ GHz, changes its sign at the gap center at $f = 1.815$ GHz, and reaches a peak at 1.82 GHz. The sign of $\eta$ is reversed when $\mu_0 H = 0.28$ T.

The strong correlation between the nonreciprocity and the gap shows that the former originated from the band structure of the superlattice. In the honeycomb lattice, the incident Rayleigh wave with wavevector $k$ near a K point of the Wigner-Seitz cell is efficiently scattered into the hybridization of three Rayleigh waves with wavevector $k$, $k - b_1$, and $k - b_2$, where $b_1$, $b_2$ are the reciprocal lattice vectors connecting the centers of Brillouin zones sharing the K point [Fig. 4(b)]. Such hybridized phonon states can induce a local strain polarization $L(r)$ (Eq. (2)). While the averaged $L$ itself vanishes [12,31], the A and B sites are surrounded by different magnetic environments, i.e. the phonons experience a periodically modulated magnon-phonon coupling, resulting in a nonreciprocity given by Eq. (1). At the same time, the scattering leads to interference and opens a gap. Both effects are characteristic to the perturbation of almost degenerate wave modes occurring at the Brillouin zone corners. Hence, the nonreciprocity is anticipated to reach a peak around the gap in the perturbed spectrum.

Based on elastic wave scattering models [36,37], phononic band dispersion, eigenmodes and $L(\mathbf{r})$ are computed, as detailed in SM Sec. 6 [31]. We show $L_A = L(\mathbf{r} = \mathbf{r}_A)$ at the high symmetry point A of each eigenmode together with the band dispersion in Fig. 4(c). The hybridization between $k$, $k - b_1$, and $k - b_2$ plane waves at $k = k_K$ opens gaps, where $k_K$ labels the wavevector at the K point. The hybridization also generates nonzero $L_A$. When $k = k_K$, the lowest band possesses nearly zero $L_A$, the middle has a positive $L_A$, and the highest a negative $L_A$. The hybridization decreases when $k$ shifts from $k_K$, and $L_A$ reduces. By time-reversal symmetry, the band structure at K' valley is the same dispersion with the opposite $L$. The dashed line shows the dispersion of the incident wave, which couples to the eigenmodes and generates the strain polarization.

Figure 4(d) shows the theoretical nonreciprocity $\eta$ (blue and red) and the $x$ component of the phonon energy flux $T$ (green) as a function of $k$, which qualitatively reproduce the tendency observed in the experiment [Fig. 4(a)]. For negative $H$ and $f$ below the gap, the incident wave is scattered into the lowest eigenmodes, which have positive $L_A$, resulting in negative nonreciprocity. Right above the gap, the incident wave couples to the highest mode, which has a large negative $L_A$, and the nonreciprocity reaches a maximum in the whole spectrum. Shifting frequency away from the gap leads to the coupling with eigenmodes with smaller $L_A$, and the smaller nonreciprocity. For positive $H$, the nonreciprocity follows the same tendency with a reversed sign. The experiment and the calculation consistently show that in the valley-selective phonon-magnon scattering, the nonreciprocity originates from the strain polarization, enabled by the plane wave hybridization in the phononic crystal.

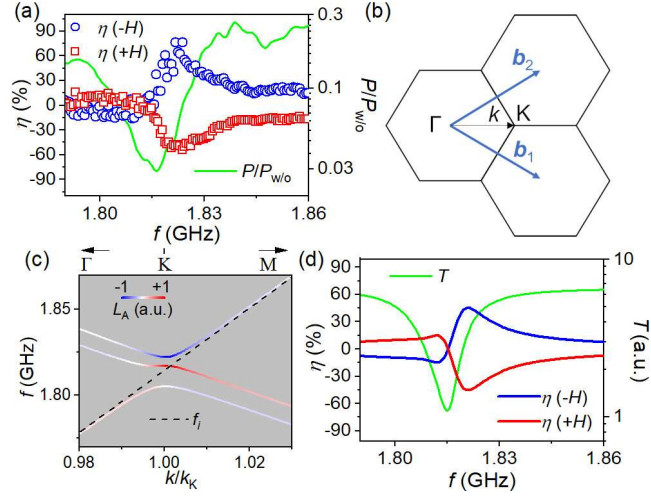

FIG. 4. (a) Nonreciprocity $\eta = (\Delta P_+ - \Delta P_-)/(\Delta P_+ + \Delta P_-)$ at the peak position as a function of SAW frequency $f$ at positive (red points) and negative (blue points) field. The green curve is the normalized transmission power $P/P_{w/o}$ as a function of $f$. (b) Schematic of the Brillouin zones sharing the K point and the relevant reciprocal lattice vectors. (c) Phonon band dispersion in the plane-wave scattering model near K point. The red, blue, and white colors represent the positive, negative, and zero strain polarization $L_A$ at site A, respectively. The wavevector at K is defined as $k_K$. The incident wave is represented by the dashed line. (d) Nonreciprocity $\eta$ (blue and red) and the $x$ component of energy flux $T$ (green) as functions of $k$ calculated from the elastic wave scattering model.

In summary, we observed valley selective phonon-magnon scattering in magnetic honeycomb superlattices, which are phononic crystals with magnon-phonon coupling. The different phonon scatterings at K and K' valleys manifest themselves as nonreciprocal SAW transmission. The nonreciprocity maintains a similar magnitude but changes sign when the superlattice symmetry is reversed. In Zigzag-like lattices without valley degrees of freedom, nonreciprocal SAW transmission becomes four times smaller, even with the inversion symmetry breaking. The valley-selective phonon-magnon scattering is maximized when the magnetic field is directed at 3° from the $z$-axis in the $xz$-plane, which can be explained by a strain polarization coupled to the magnetization through the conventional magnetoelastic interaction. The nonreciprocity spectrum shows a peak near the transmission gap, matching with the distribution of the strain polarization in the phonon bands. Different from the edge modes, our approach provides a way to apply valley phonon in the bulk transport regime. It demonstrates the periodically modulated magnon-phonon coupling as a new tool for controlling the valley phonon transport.

**Acknowledgments:**

We thank Yoichi Nii, Konstantin Bliokh, Lei Han, Huiping Xu, and Joseph Barker for fruitful discussions and helpful comments. This work is supported by JSPS KAKENHI 19H05629. KY acknowledges support from JST PRESTO Grant No. JPMJPR20LB,

Japan and JSPS KAKENHI (No. 21K13886), and YO from The LANEF Chair of Excellence, QSPIN project, at University Grenoble Alpes. SM is financially supported by JST CREST (No.JPMJCR19J4, No.JPMJCR1874, and JPMJCR20C) and JSPS KAKENHI (No.20H01865) from MEXT Japan. LL would like to thank the support from JSPS through "Research program for Young Scientists" (no. 23KJ0778). YH acknowledges support from RIKEN Junior Research Associate Program.

*Pumping with Coherent Elastic Waves*, Phys. Rev. Lett. **108**, 176601 (2012).

[36] M. S. Kushwaha, P. Halevi, G. Martínez, L. Dobrzynski, and B. Djafari-Rouhani, *Theory of Acoustic Band Structure of Periodic Elastic Composites*, Phys. Rev. B **49**, 2313 (1994).

[37] E. N. Economou and M. Sigalas, *Stop Bands for Elastic Waves in Periodic Composite Materials*, J. Acoust. Soc. Am. **94**, 1734 (1994).

[38] M. Hu and F. Li Duan, *Design, Fabrication and Characterization of SAW Devices on LiNbO3 Bulk and ZnO Thin Film Substrates*, Solid State Electron. **150**, 28 (2018).

[39] V. S. Bhat, F. Heimbach, I. Stasinopoulos, and D. Grundler, *Magnetization Dynamics of Topological Defects and the Spin Solid in a Kagome Artificial Spin Ice*, Phys. Rev. B **93**, 140401 (2016).

[40] Q. Li, S. Xiong, L. Chen, K. Zhou, R. Xiang, H. Li, Z. Gao, R. Liu, and Y. Du, *Spin-Wave Dynamics in an Artificial Kagome Spin Ice*, Chin. Phys. Lett. **38**, 047501 (2021).

[41] R. Su et al., *Wideband and Low-Loss Surface Acoustic Wave Filter Based on 15° YX-LiNbO/SiO/Si Structure*, IEEE Electron Device Lett. **42**, 438 (2021).

[42] K. Yamamoto, M. Xu, J. Puebla, Y. Otani, and S. Maekawa, *Interaction between Surface Acoustic Waves and Spin Waves in a Ferromagnetic Thin Film*, J Magn. Magn. Mater. **545**, 168672 (2022).

## Supplemental Material: Valley-Selective Phonon-Magnon Scattering in Magnetoelastic Superlattices

1. Transmission spectra and sample layout

The transmission spectra of delay line devices with and without superlattice inside are shown in Fig. S1. The delay line device with superlattice has a dip in the transmission, corresponding to the gap induced by the superlattice. Based on the transmission spectra, we can get the transmission power $P_0(S12^2+S21^2)/2$, where $P_0 = 10$ mW is the applied power, labeled as $P$ for device with superlattice and $P_{w/o}$ for device without superlattice.

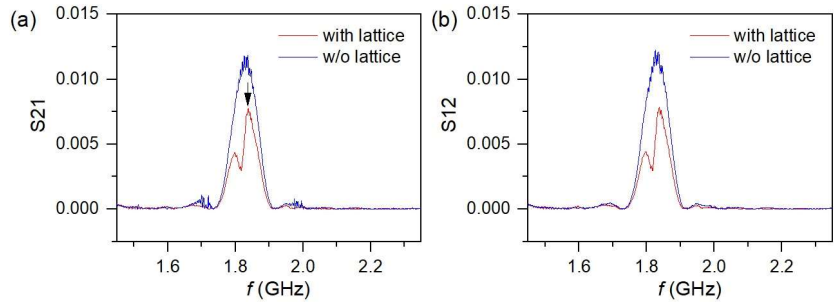

FIG. S1. S21 (a) and S12 (b) transmission spectra of SAW delay line device without superlattice (blue) and with superlattice (red).

The magnetic field dependence of the transmission in Fig. 2 and Fig. 3 is measured at the peak frequency $1.836 \pm 0.001$ GHz, labeled by arrow in Fig. S1(a).

Figure S2(a) shows the detail of the geometry of the lattice structure and Fig. S1(b) shows the geometry of the superlattice and IDTs. The IDTs are made of 35 nm Al film.

The separation between port 1 and port 2 IDTs is 300 μm. The superlattice and IDTs were fabricated by e-beam lithography followed by e-beam evaporation.

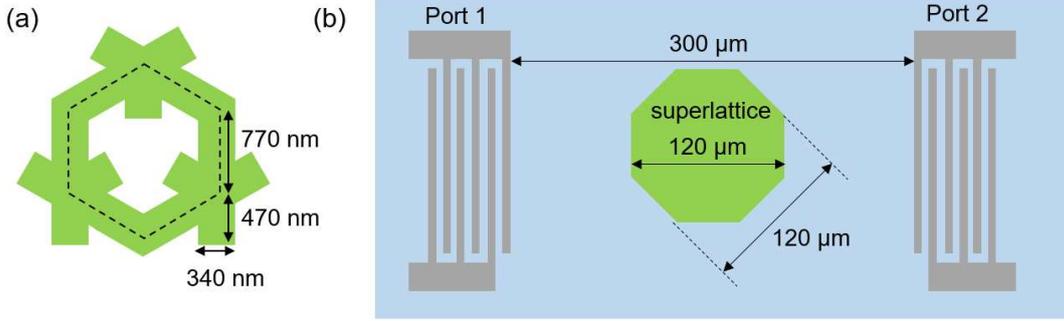

FIG. S2. (a) Geometry of the lattice structure. (b) Geometry of the SAW delay line device with superlattice inside.

## 2. Influence of the substrate orientation

The 128° Y-LiNbO$_3$ substrate has large piezoelectric effect, allowing us to generate phonons efficiently, but the effect is anisotropic in the substrate plane. We measure the peak transmission frequency of SAW delay lines with wavelength 2 μm oriented at different directions on the 128° Y-LiNbO$_3$ substrate, as shown in Fig. S3. Since the velocity $c = \lambda f$ and we keep $\lambda = 2$ μm, the variation of the frequency represents the variation of the velocity $c$. We define $\phi_k$ as the angle between the SAW wavevector and the Y' axis of the substrate. While velocity varies significantly in the whole range, the velocities at 0°, 60° and 120° are almost the same. The frequency difference of these angles is of the order of 10 MHz, one order smaller than the phonon linewidth given in Fig. S1. The same orientation dependence of the Rayleigh wave velocity is shown in Fig. 12(a) in [1]. Thus, when the superlattice has K direction along the Y' axis, the three plane waves with wavevectors at 0°, 60° and 120° are almost degenerate and can hybridize efficiently. We hence select the Y' axis as the SAW propagation direction in this experiment.

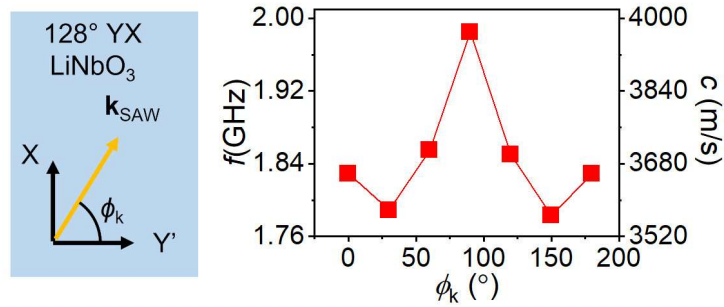

FIG. S3. Orientation dependence of the Rayleigh wave velocity on 128° YX-LiNbO$_3$ substrate.

## 3. In-plane scan and low field absorption

The results of in-plane scan are shown in Fig. S4. We measured the transmission when the magnetic field is rotated by angle $\phi$ from the SAW wavevector [Fig. S4(a)].

For $\phi = 90°$ [Fig. S4(b)], $\phi = 45°$ (c), and $\phi = 0°$ (d), absorption always exists, different from usual unpatterned Ni films that have largest absorption at 45° but no absorption at 0° and 90° [2]. Especially, the absorption exists at zero field for all three angles and shows butterfly-like hysteresis at $\phi = 0°$. No nonreciprocity is observed in the in-plane scanning. These behaviors are similar to the low-field absorption peak in the out-of-plane scanning. Also, the zero-field state in the out-of-plane scanning should be similar to the zero field states in the in-plane scanning, since they are all thought to be multidomain states. Hence, the low-field absorption peak may have the same origin as the in-plane absorption peaks. One possible explanation is that the SAW is absorbed by magnetic resonances under the internal, possibly spatially inhomogeneous, effective field, which is due to the shape anisotropy of the superlattice structure [3,4].

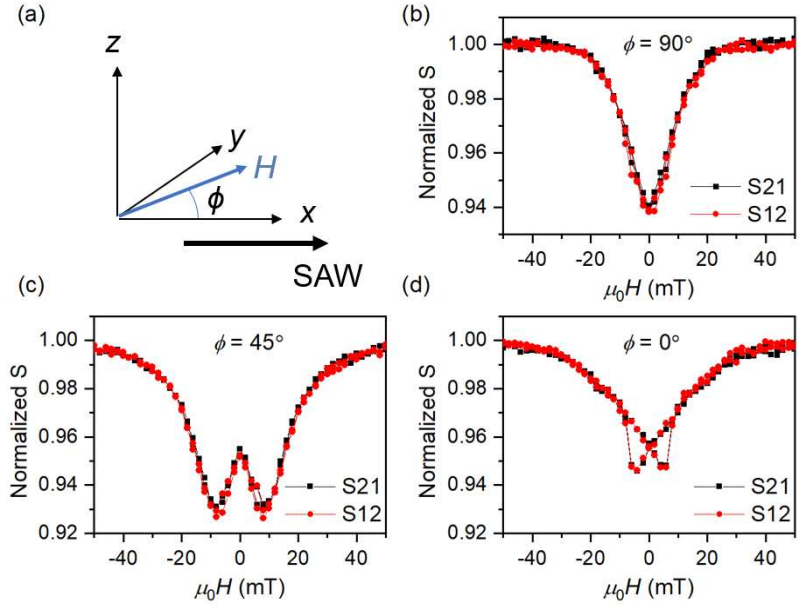

FIG. S4. (a) Setup for the in-plane transmission measurement. $\phi$ is the angle between the magnetic field and the SAW wavevector. (b-d) Normalized transmission as a function of magnetic field for $\phi = 90°$ (b), $\phi = 45°$ (c), and $\phi = 0°$ (d).

4. Derivation of the theoretical nonreciprocal power

Let $\boldsymbol{u}(\boldsymbol{r})$ be the displacement vector and $\boldsymbol{M}(\boldsymbol{r}) = M_S \boldsymbol{n}(\boldsymbol{r})$ be a spatial distribution of magnetization, which model the deformation of the whole sample structure and the magnetic moments in the patterned Ni layer in our experiment respectively. Since the Ni layer is thought to be polycrystalline, and considering the large number of free parameters of general anisotropic magneto-elastic coupling, we assume an approximate rotational symmetry and base our analysis on the free energy density

$$F_{me} = b \sum_{i,j=x,y,z} \varepsilon_{ij} n_i n_j \, , \text{(S1)}$$

where $\varepsilon_{ij} = (\partial_j u_i + \partial_j u_i)/2$ is the linear strain tensor. The effective magnetic field induced by the strain $\varepsilon_{ij}$ is defined to be

$$h_i^{\text{eff}} = -\frac{2b}{\mu_0 M_S} \sum_{j=x,y,z} \varepsilon_{ij} n_j \, , \text{(S2)}$$

where $\mu_0$ is the vacuum permeability. Since $\varepsilon_{ij}$ is small, we set $n_j$ on the right-hand-side to be their ground state values. Aligning the conventions with the experiment, we fix the $x$ axis along the in-plane component of $\mathbf{n}$ and denote the angle between $\mathbf{n}$ and $z$ axis by $\theta$, i.e. $n_x = \sin\theta$, $n_y = 0$, $n_z = \cos\theta$. Only the components of $\mathbf{h}^{\text{eff}}$ in the plane perpendicular to the ground state $\mathbf{n}$ contribute to the linear magnetization dynamics, which are given by

$$h_{x'}^{\text{eff}} \equiv h_x^{\text{eff}} \cos\theta - h_z^{\text{eff}} \sin\theta$$
$$= -\frac{2b}{\mu_0 M_S}[(\varepsilon_{xx} - \varepsilon_{zz})\cos\theta\sin\theta + \varepsilon_{zx}\cos 2\theta], \quad \text{(S3)}$$

$$h_y^{\text{eff}} = -\frac{2b}{\mu_0 M_S}(\varepsilon_{xy}\sin\theta + \varepsilon_{yz}\cos\theta). \quad \text{(S4)}$$

The linearized Landau-Lifshitz-Gilbert equation takes the form

$$\left\{F^{(2)} - \frac{iM_S}{\gamma}\begin{pmatrix} \alpha\omega & -\omega \\ \omega & \alpha\omega \end{pmatrix}\right\}\begin{pmatrix} n_{x'} \\ n_y \end{pmatrix} = \mu_0 M_S \begin{pmatrix} h_{x'}^{\text{eff}} \\ h_y^{\text{eff}} \end{pmatrix}, \quad \text{(S5)}$$

where Fourier transform in time has been applied, $\gamma$ is the gyromagnetic ratio, $\alpha$ is the Gilbert damping, and $F^{(2)}$ is a self-adjoint two-by-two matrix operator determined by the second order perturbative expansion of the magnetic part of the free energy density. For the simplest case of static magnetic field $H$ parallel to the ground state $\mathbf{n}$,

$$F^{(2)} = \mu_0 M_S H \begin{pmatrix} 1 & 0 \\ 0 & 1 \end{pmatrix}. \quad \text{(S6)}$$

Although in general $F^{(2)}$ may contain spatial derivatives as with exchange interactions, non-local contributions as for dipole-dipole interactions, and terms not proportional to the identity matrix arising from any anisotropy in the system, we ignore them all for simplicity since the purpose of this section is a qualitative demonstration of non-reciprocity. It would be justified for paramagnetic materials, but we shall discuss more realistic situations in the following sections. One may then solve the equation locally to obtain

$$\begin{pmatrix} n_{x'} \\ n_y \end{pmatrix} = \frac{\gamma\mu_0}{\omega_F^2 - \omega^2 - i\alpha\omega_F\omega}\begin{pmatrix} (\omega_F - i\alpha\omega)h_{x'}^{\text{eff}} - i\omega h_y^{\text{eff}} \\ i\{\omega h_{x'}^{\text{eff}} - i(\omega_F - i\alpha\omega)h_y^{\text{eff}}\} \end{pmatrix}, \quad \text{(S7)}$$

where second order contributions from $\alpha$ have been dropped, and $\omega_F = \gamma\mu_0 H$. To leading order in the magneto-elastic coupling, the phonon absorption should be proportional to the work per time per unit volume $\Delta P$ done by $\mathbf{h}^{\text{eff}}$, i.e.

$$\Delta P \equiv \text{Re}\left\{-i\omega\mu_0 M_S\left(n_{x'}\overline{h_{x'}^{\text{eff}}} + n_y \overline{h_y^{\text{eff}}}\right)\right\}$$
$$= \alpha\omega^2\gamma\mu_0^2 M_S \frac{\omega^2\left(|h_{x'}^{\text{eff}}|^2 + |h_y^{\text{eff}}|^2\right) - i\omega\omega_F\left(h_y^{\text{eff}}\overline{h_{x'}^{\text{eff}}} - h_{x'}^{\text{eff}}\overline{h_y^{\text{eff}}}\right)}{\left(\omega_F^2 - \omega^2\right)^2 + \alpha^2\omega_F^2\omega^2}, \quad \text{(S8)}$$

where the overbar denotes complex conjugation. While the first term in the numerator always gives a positive contribution, the second term can be either positive or negative, depending on whether the polarization of the effective magnetic field is right- or left-handed circular. Near the resonance $\omega^2 \sim \omega_F^2$, a fully circularly polarized field couples either twice as strongly as linear ones or almost completely decouples, which is the well-recognized chirality of ferromagnetic resonance. We denote this polarization dependent term $\Delta P_{\text{chiral}}$.

Let us consider the effective magnetic field generated by the surface acoustic waves in the patterned region. For the Rayleigh type surface acoustic waves scattered by periodic structures patterned on the surface, quantitative description is too complicated while we do not aim for any quantitative agreement with the experiment, so we consider bulk phonons propagating in a periodic structure that is spatially homogeneous in $z$-direction, reducing the problem to two-dimensional one. A qualitative description of the experiment can be provided in this way since for the Rayleigh-type plane waves on 128° Y-LiNbO$_3$, the longitudinal strain is predominant, and the strains involving $z$ derivations are smaller. For the two-dimensional waves, $\varepsilon_{xz} = \varepsilon_{zy} = \varepsilon_{zz} = 0$ so that

$$\Delta P_{\text{chiral}} = \frac{\alpha \gamma b^2 \omega^2 \omega_F}{M_S} \frac{L \sin 2\theta \sin \theta}{\left(\omega_F^2 - \omega^2\right)^2 + \alpha^2 \omega_F^2 \omega^2}, \quad (S9)$$

where

$$L = -2i\omega(\overline{\varepsilon_{xx}}\varepsilon_{xy} - \overline{\varepsilon_{xy}}\varepsilon_{xx}) \doteq \left(\varepsilon_{xx}\, d\varepsilon_{xy}/dt - \varepsilon_{xy}\, d\varepsilon_{xx}/dt\right), \quad (S10)$$

with $\doteq$ indicating the equality upon Fourier transform. Therefore, the chirality of the acoustic waves is encoded in $L$ in the sense that it decides the polarization of the effective magnetic field, and its contribution to absorption depends on the out-of-plane angle as $\sin 2\theta \sin \theta$.

As a closing remark, we clarify the relation between $\Delta P_{\text{chiral}}$ and nonreciprocity. By definition, $L$ is odd in $\omega$, and so is $\Delta P_{\text{chiral}}$. For classical waves, the sign of $\omega$ on its own does not have any physical significance and is just a matter of convention. But when the Fourier transform in space is also introduced, the *relative* sign between $\omega$ and $k$ has the physical meaning of propagation direction. Due to the reality of classical field variables, any physical observable should contain only even powers in $\omega$, $k$, and any dependence on the direction of propagation should involve odd powers of both $\omega$ and $k$ (therefore even in $\omega$, $k$). The odd power in $k$ leads to the nonreciprocity, so that the nonreciprocal absorption $\Delta P_n$ in the experiment can be modeled by $\Delta P_{\text{chiral}}$.

## 5. Estimation of the magnetization tilting angle

We measured the magnetic field at which the Ni magnetization in our superlattice saturates using the anomalous Hall effect. The superlattice is fabricated in a Hall bar device with channel dimension 100 μm × 50 μm. Current $I$ = 1 mA is applied, and Hall voltage $V_H$ is measured while scanning magnetic field along the out-of-plane direction. As shown in Fig.S5, the anomalous Hall resistance R = $V_H / I$ becomes constant when the magnetic field is above 0.35 T, revealing that the magnetization is saturated out-of-plane. We then use $M_S \approx 2.8 \times 10^5$ A/m for the estimation of the magnetization tilting angle.

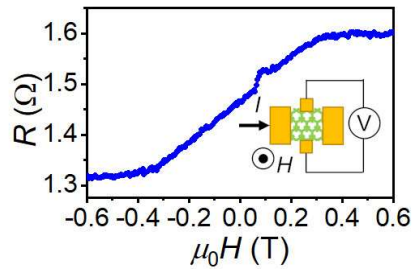

FIG. S5. Anomalous Hall resistance as a function of out-of-plane magnetic field in a superlattice device. Inset shows the measurement geometry.

The magnetization tilting angle $\theta$ is calculated using a macrospin model displayed in Fig. S6(a). The free energy is

$$F_m = -\mu_0 M_S H(\cos\theta_H \cos\theta + \sin\theta_H \sin\theta \cos\phi) + K_{an}\cos^2\theta, \quad (S11)$$

where $H$ is the applied field, $\theta_H$ is the magnetic field tilting angle, $\phi$ is the in-plane angle of the magnetization, the saturated magnetization $M_S \approx 2.8\times10^5$ A/m, and anisotropy $K_{an} \approx \mu_0 M_S^2/2$ mainly arising from the demagnetization field. $\phi$ is always 0° in the equilibrium state. For each $\theta_H$, $\theta$ as a function of $H$ is calculated by minimizing $F_m$, as plotted in Fig. S6(b). Under small field, the magnetization is in-plane due to the demagnetization field, corresponding to $\theta \sim 90°$. When the field increases, the magnetization rotates towards the out-of-plane direction, and $\theta$ decreases. Smaller $\theta_H$ rotates the magnetization faster.

The FMR frequency can be calculated from the Landau-Lifshitz equation. For the present consideration, one may drop the damping and effective fields from Eq. (S5) to obtain

$$iM_S\omega\begin{pmatrix}n_{x'}\\n_y\end{pmatrix} = \gamma\begin{pmatrix}0 & \frac{\partial^2 F_m}{\partial n_y^2}\\ -\frac{\partial^2 F_m}{\partial n_{x'}^2} & 0\end{pmatrix}\begin{pmatrix}n_{x'}\\n_y\end{pmatrix} = \gamma\begin{pmatrix}0 & \frac{1}{\sin^2\theta}\frac{\partial^2 F_m}{\partial \phi^2}\\ -\frac{\partial^2 F_m}{\partial \theta^2} & 0\end{pmatrix}\begin{pmatrix}n_{x'}\\n_y\end{pmatrix},(S12)$$

where $n_{x'} = \Delta\theta$ and $n_y = \sin\theta\Delta\phi$, $\omega$ is the angular frequency of the mode, and the diagonal terms are zero because $\partial^2 F_m/\partial\theta\partial\phi = \partial^2 F_m/\partial\phi\partial\theta = 0$ for $\phi = 0°$. Taking $\gamma/2\pi = 28$GHz/T, diagonalizing Eq. (S12) with the $\theta$ given by minimizing Eq. (S11), we get the FMR eigenfrequency $\omega_{FMR}$, and plot $f_{FMR} = \omega_{FMR}/2\pi$ in Fig. S5(c). Then the resonance condition $f_{FMR} = f_{SAW}$ is found by taking the crossing points of the FMR curve and the horizontal line $f = f_{SAW}$. Each crossing point provides one magnetic field, which corresponds to one magnetization angle $\theta$ in Fig. S5(b). In this way, we get the evolution of $\theta$ at the resonance condition as a function of $\theta_H$, which is used to calculate the $\sin2\theta\sin\theta$ vs. $\theta_H$ curve in Fig. 3(c). We model the Ni magnetization under resonance condition $f_{FMR} = f_{SAW}$ by the paramagnetic moments under the same resonance condition and with the same tilting angle $\theta$, i.e., replacing the $\omega_F$ in Eqs. (S8) and (S9) with $\omega_{FMR}$ obtained from Eq. (S12). Under this approximation, $\sin2\theta\sin\theta$ vs. $\theta_H$ curve may be interpreted as the $\Delta P_{chiral}$ vs. $\theta_H$ curve, providing the $\theta_H$ dependence of the nonreciprocal power.

The $f_{FMR}$ curves for $\theta_H = 3°$ and $\theta_H = 0°$ are highlighted in green and brown colors in Fig. S6(c), respectively. The curve for $\theta_H = 3°$ crosses the line $f = f_{SAW}$ at around 0.3 T, matching with the resonance peak at $\sim 0.3$ T in the transmission spectra. The $f_{FMR}$ curve is almost flat in this regime, so that the SAW nearly meets the resonance condition in a wide magnetic field range, leading to a broad absorption peak in the transmission spectra. At $\theta_H = 0°$, the $f_{FMR}$ curve is given by the out-of-plane FMR condition $\omega = \gamma\mu_0(H - M_S)$, which is a straight line crossing $f = f_{SAW}$ at $\sim 0.43$ T. To examine this, we show the zoom-in plot of the normalized S around $\pm 0.43$ T at $\theta_H = 0°$ in Fig. S6(d). Indeed, we can see small absorption peaks at around $\pm 0.43$ T. The magnetization ground state is fully out-of-plane under this condition, so that it only couples to the out-of-plane shear strain components [5], which are small in Rayleigh waves in thin films [6]. Hence, the absorption amplitude is rather small compared with the peaks at $\theta_H = 3°$.

Restricted by the device bandwidth and the signal amplitude, we cannot map the wide frequency range absorption spectra using the superlattice devices directly. Alternatively, we fabricated X-oriented delay line SAW devices without superlattice, as shown in Fig. S6(e). The large electromechanical coupling coefficient in such devices leads to large signal and large bandwidth [7], allowing us to map the wide frequency range absorption spectra. In this measurement, we use delay lines with IDT wavelength 2.28 μm, 1.8 μm and 1.28 μm for spectra from 1.3 GHz to 1.95 GHz, from 1.95 GHz to 2.65 GHz, and from 2.65 GHz to 3.3 GHz, respectively. Figure S6(f) plots the normalized S12 at $\theta_H = 3°$, where dark color represents absorption. The absorption pattern well follows the FMR curve at $\theta_H = 3°$ calculated in Fig. S6(c). Figure S6(g) shows the results at $\theta_H = 0°$, which are close to straight lines. The results are also close to the calculation shown in Fig. S6(c). Though, the SAW couples to magnons with finite wavelength, whose frequency is different from the FMR, and the angle resolution of the used vector electromagnet is ~ 1°, so that the experimental curves do not, and should not be expected to, agree quantitatively with the toy model calculation.

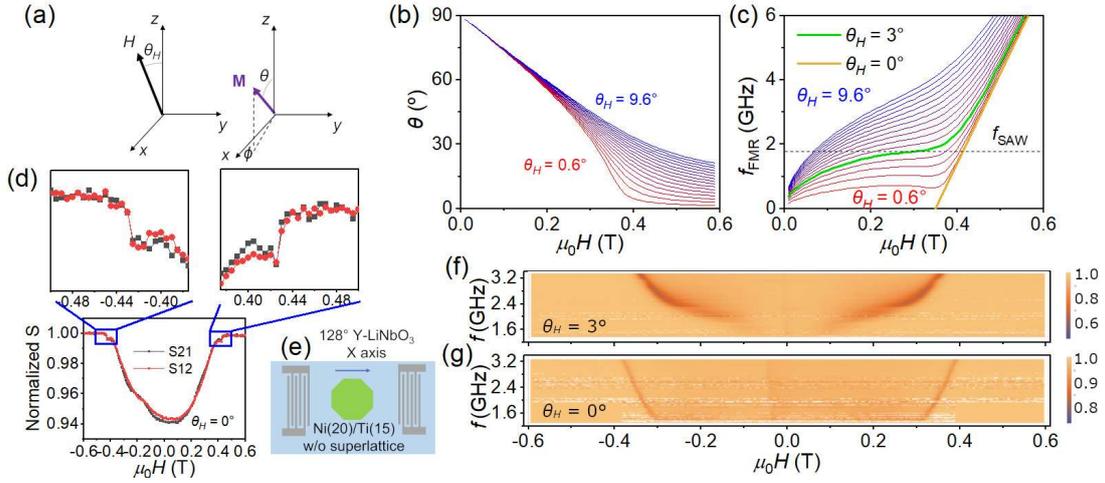

FIG. S6. (a) Schematic of the marcospin model. The magnetic field is in the $xz$ plane, with an angle $\theta_H$ tilted from the $z$ axis, $\theta$ is the angle between the magnetization $M$ and the $z$ axis, and $\phi$ is the angle between the in-plane projection of $M$ and the $x$ axis. (b) Magnetization tilting angle $\theta$ as a function of magnetic field $H$ with different magnetic field tilting angle $\theta_H$. (c) FMR frequency $f_{FMR}$ as a function of magnetic field $H$ with different magnetic field tilting angle $\theta_H$. (d) Zoom-in plot of the transmission spectra of the $\theta_H = 0°$ data for the superlattice device. At ± 0.43 T, small absorption peaks are present. (e) Layout of the IDT delay line devices for spectrum mapping. Ni(20)/Ti(15) film was deposited without superlattice structure, and the SAW is sent along the X axis of the 128° Y-LiNbO$_3$ substrate. (f-g) Normalized S12 spectra from 1.3 GHz to 3.3 GHz under magnetic field from – 0.6 T to 0.6 T at $\theta_H = 3°$ (f) and $\theta_H = 0°$ (g). The color bar represents the transmission parameter S12 normalized by a reference S12 value at 0.6 T. The plot range is 0.5-1 for $\theta_H = 3°$ (f) and 0.75-1 for $\theta_H = 0°$ (g).

6. Elastic wave scattering model and theoretical nonreciprocity

Here, we demonstrate that the periodic patterns used in the experiment may induce nonzero $\varepsilon_{xy}$ whose phase differs from $\varepsilon_{xx}$ of the incident Rayleigh wave. As stated in Sec. S4, we consider bulk phonons propagating in a periodic structure that is spatially homogeneous in $z$-direction. According to Ref. [8,9], the problem can be modelled by

$$\omega^2 u_{k+G} = \sum_{G',G''} \rho^{-1}_{G-G''}[\lambda_{G''-G'}(k+G'')(k+G') \cdot u_{k+G'} \mu_{G''-G'} \\ +\{(k+G')(k+G'') \cdot u_{k+G'} + (k+G') \cdot (k+G'')u_{k+G'}\}]. \quad (S13)$$

Here $\mathbf{k}$ is the wavevector in the first Brillouin zone, $\mathbf{G} = m_1 \mathbf{b}_1 + m_2 \mathbf{b}_2$ is the reciprocal lattice vector, $\rho^{-1}{}_\mathbf{G}$, $\lambda_\mathbf{G}$, $\mu_\mathbf{G}$ are Fourier coefficients of the periodic mass density and Lamé constants respectively, and $\mathbf{u}_{\mathbf{k}+\mathbf{G}}$ is the Fourier component of the displacement vector with wavevector $\mathbf{k} + \mathbf{G}$. We focus on the waves spatially uniform in $z$ direction so that $\mathbf{k} = (k_x, k_y, 0)^T$. In this case, the $z$ component of $\mathbf{u}_{\mathbf{k}+\mathbf{G}}$ decouples from the rest and we further restrict our attention to the in-plane components.

We are interested in the situation where the spatial modulation is very weak so that the problem can be treated as a perturbation to the spatially homogeneous problem. Namely, we assume $|\rho_a^{-1}| \gg |\rho_{\mathbf{G}\neq 0}^{-1}|$, where $\rho_a^{-1}$ is the averaged value of $\rho^{-1}$, and similarly for $\lambda$, $\mu$. We use subscript $a$ to distinguish from the vacuum permeability $\mu_0$. In the unperturbed system, the eigenmodes are the usual longitudinal and transverse plane waves:

$$\mathbf{u}_{\mathbf{k}+\mathbf{G}}^P = \frac{\mathbf{k}+\mathbf{G}}{|\mathbf{k}+\mathbf{G}|}, \mathbf{u}_{\mathbf{k}+\mathbf{G}}^S = \hat{z} \times \frac{\mathbf{k}+\mathbf{G}}{|\mathbf{k}+\mathbf{G}|}, \quad \text{(S14)}$$

with the eigenfrequencies $\omega_P^2 = c_P^2 |\mathbf{k}+\mathbf{G}|^2$, $\omega_S^2 = c_S^2 |\mathbf{k}+\mathbf{G}|^2$ where $c_P^2 = \rho_a^{-1}(\lambda_a + 2\mu_a)$, $c_S^2 = \rho_a^{-1}\mu_a$. Note that the homogeneous components of the Fourier coefficients $\rho_a^{-1}$, $\lambda_a$, $\mu_a$ describe the average values of $\rho^{-1}$, $\lambda$, $\mu$, so that they are affected by the presence of the surface patterning albeit in an inessential way. The polarization characteristic of the Rayleigh wave is mainly that of longitudinal wave $\mathbf{u}_{\mathbf{k}+\mathbf{G}}^P$ whose perturbation we are going to study.

The periodic modulation will generate mixing of $\mathbf{u}_{\mathbf{k}+\mathbf{G}}^P$ with $\mathbf{u}_{\mathbf{k}+\mathbf{G}+\mathbf{G}'}^{P,S}$. The mixing coefficient is usually inversely proportional to $c_P^2 |\mathbf{k}+\mathbf{G}|^2 - c_{P,S}^2|\mathbf{k}+\mathbf{G}+\mathbf{G}'|^2$. In our setup, $\mathbf{b}_1$, $\mathbf{b}_2$ will be of similar order of magnitude to the incident $\mathbf{k} + \mathbf{G}$ so that the mixing will be suppressed by $\sim 2\delta c_{P,S}/c_{P,S}$ where $\delta c_{P,S}$ is the difference in sound velocities between the unpatterned and patterned areas. However, it is possible to have $c_P |\mathbf{k}+\mathbf{G}| = c_{P,S} |\mathbf{k}+\mathbf{G}+\mathbf{G}'|$ at some special points in the Brillouin zone for some special values of $\mathbf{G}'$. Near those points, the mixing will be of order unity and the influence of the periodic structure will be enhanced. This almost degenerate situation is always realized near the edges of the Wigner-Seitz cell, i.e., $c_P |\mathbf{k}| \approx c_P |\mathbf{k}+\mathbf{G}'|$ for some $\mathbf{G}'$. Although there can be other accidental degeneracies, they are not expected to appear generically unless some other aspects are taken into account and we ignore them in this study. The leading order perturbation under this approximation is

$$\omega^2 \mathbf{u}_{\mathbf{k}+\mathbf{G}} \approx c_P \left| \mathbf{k}+\mathbf{G} \right|^2 \mathbf{u}_{\mathbf{k}+\mathbf{G}} + \sum_{\mathbf{G}'\in\Omega} \Big[ \rho_a^{-1} \lambda_{\mathbf{G}-\mathbf{G}'}(\mathbf{k}+\mathbf{G})(\mathbf{k}+\mathbf{G}') \cdot \mathbf{u}_{\mathbf{k}+\mathbf{G}'}$$
$$+ \rho_a^{-1} \mu_{\mathbf{G}-\mathbf{G}'} \{(\mathbf{k}+\mathbf{G}')(\mathbf{k}+\mathbf{G}) \cdot \mathbf{u}_{\mathbf{k}+\mathbf{G}'} + (\mathbf{k}+\mathbf{G}') \cdot (\mathbf{k}+\mathbf{G}) \mathbf{u}_{\mathbf{k}+\mathbf{G}'}\} \quad \text{(S15)}$$

where $\Omega$ denotes the set of reciprocal vectors $\mathbf{G}'$ for which $|\mathbf{k}+\mathbf{G}| \approx |\mathbf{k}+\mathbf{G}'|$ and $\mathbf{G} \neq \mathbf{G}'$. Note that the $\rho_{\mathbf{G}\neq 0}^{-1}$ terms are ignored, as the square time-evolution operator of the acoustic waves with inhomogeneous density is not self-adjoint with respect to the Euclidean metric [10], making the problem non-Hermitian when written in the matrix form with these terms. As a qualitative modelling, we set $\rho$ as a constant.

The actual set $\Omega$ depends on $\mathbf{G}$, so it is more convenient to define the set of vectors $\mathbf{k}_n$, $n = 0, 1, \ldots, N$ where $N$ is the number of elements in $\Omega$, $\mathbf{k}_n - \mathbf{k}_{n'}$ is a reciprocal vector, and $|\mathbf{k}_n| \approx |\mathbf{k}_{n'}|$ for every $n$ and $n'$. The vectors $\mathbf{k}_n$ should be symmetrically distributed near a circle centered on $\Gamma$-point. Finally, we use the ansatz $\mathbf{u}_{\mathbf{k}_n} = A_n \mathbf{u}_{\mathbf{k}_n}^P$ since the transverse modes are not expected to be degenerate, and project the equations onto $\mathbf{u}_{\mathbf{k}_n}^P$. In this way, one should obtain a well-posed $N+1$ by $N+1$ eigenvalue problem for the column vector $\{A_n\}$.

6.1 K-point in honeycomb lattice

We consider a honeycomb lattice as shown in Fig. S6(a). The coordinates of site A and B are $r_A = (0, 0, 0)$, $r_B = (0, a, 0)$, respectively, where $a$ is the length of a side of the hexagon (not the period itself). Let $a_1$, $a_2$ be the real space basis vectors of the honeycomb lattice. We fix the convention

$$a_1 = \sqrt{3}a \begin{pmatrix} 1/2 \\ -\sqrt{3}/2 \\ 0 \end{pmatrix}, a_2 = \sqrt{3}a \begin{pmatrix} 1/2 \\ \sqrt{3}/2 \\ 0 \end{pmatrix}. \quad (S16)$$

The corresponding reciprocal lattice vectors are given by

$$b_1 = \frac{4\pi}{3a} \begin{pmatrix} \sqrt{3}/2 \\ -1/2 \\ 0 \end{pmatrix}, b_2 = \frac{4\pi}{3a} \begin{pmatrix} \sqrt{3}/2 \\ 1/2 \\ 0 \end{pmatrix}. \quad (S17)$$

The Brillouin zone is shown in Fig. S6(b).

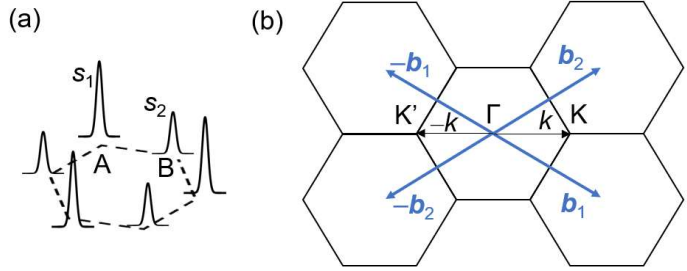

FIG. S6. (a) Schematic of the honeycomb lattice constructed by delta-scattering functions. (b) Brillion zone and the corresponding reciprocal vectors in the model.

We define the K- and K'-point to be $k_K = (b_1 + b_2)/3$, $k_{K'} = -(b_1 + b_2)/3$, and $|k_K| = |k_{K'}| = k_K$. For this choice $\Omega = \{-b_1, -b_2\}$, $\{b_1, b_2\}$ respectively. We therefore study the coupling of three longitudinal modes with wavevectors $k_0 = k_K + q, k_1 = k_0 \mp b_1, k_2 = k_0 \mp b_2, |q| \ll 2\pi/a$. The equations read

$$\omega^2 A_0 = c_P^2 k_0^2 A_0 + \rho_a^{-1}\left(\lambda_{\pm b_1} + 2\mu_{\pm b_1}\cos^2\phi_{01}\right)k_0 k_1 A_1 + \rho_a^{-1}\left(\lambda_{\pm b_2} + 2\mu_{\pm b_2}\cos^2\phi_{02}\right)k_0 k_2 A_2, \quad (S18)$$

$$\omega^2 A_1 = c_P^2 k_1^2 A_1 + \rho_a^{-1}\left(\lambda_{\mp b_1} + 2\mu_{\mp b_1}\cos^2\phi_{01}\right)k_0 k_1 A_0 + \rho_a^{-1}\left(\lambda_{\mp b_1 \pm b_2} + 2\mu_{\mp b_1 \pm b_2}\cos^2\phi_{12}\right)k_1 k_2 A_2, \quad (S19)$$

$$\omega^2 A_2 = c_P^2 k_2^2 A_2 + \rho_a^{-1}\left(\lambda_{\mp b_2} + 2\mu_{\mp b_2}\cos^2\phi_{02}\right)k_0 k_2 A_0 + \rho_a^{-1}\left(\lambda_{\pm b_1 \mp b_2} + 2\mu_{\pm b_1 \mp b_2}\cos^2\phi_{12}\right)k_1 k_2 A_2, \quad (S20)$$

where $k_n = k_n(\cos\phi_n, \sin\phi_n, 0)^T$, $n = 0,1,2$ and $\phi_{nn'} = \phi_n - \phi_{n'}$. The following should hold: $k_0 \approx k_1 \approx k_2$, $\phi_{01} \approx -\phi_{02} \approx \phi_{12} \approx 2\pi/3$.

For simplicity, we take $\lambda(r)/\lambda_a = \mu(r)/\mu_a = f(r)$, where $f(r)$ is a periodic function on the lattice, given by

$$f(r) = S_0 \sum_{R_i}\{s_1\delta(r - R_i - r_A) + s_2\delta(r - R_i - r_B) - (s_1 + s_2)/S_0\}, \quad (S21)$$

where $s_1$ and $s_2$ are dimensionless parameters labeling the strength of the point-source scattering on A- and B-sites, respectively, $S_0$ is the unit cell area, and $R_i$ is summed over the Bravais lattice sites. Then it is easy to obtain the Fourier coefficients $f_{b_1} = s_1 + s_2 t_1, f_{b_2} = f_{b_1 - b_2} = s_1 + s_2 t_2,$ where $t_1 = \exp(2\pi i/3)$, $t_2 = \exp(-2\pi i/3)$, and $f_{-G} = \overline{f_G}$.

Once the eigenvalue problem is solved, one should obtain three eigenvectors $\{A_0^m, A_1^m, A_2^m\}$, $m = 1, 2, 3$ belonging to the eigenvalues $\omega_m^2$. Each of them represents a physical acoustic wave given by

$$\boldsymbol{u} \equiv \mathrm{Re}\left[\boldsymbol{v} e^{-i\omega_m t}\right] = \mathrm{Re}\left[e^{-i\omega_m t}\left(A_0^m e^{i\mathbf{k}_0 \cdot \mathbf{r}} \frac{\mathbf{k}_0}{k_0} + A_1^m e^{i\mathbf{k}_1 \cdot \mathbf{r}} \frac{\mathbf{k}_1}{k_1} + A_2^m e^{i\mathbf{k}_2 \cdot \mathbf{r}} \frac{\mathbf{k}_2}{k_2}\right)\right]. \quad (S21)$$

The associated strain tensor components read

$$\varepsilon_{xx} = -\mathrm{Im}\left[e^{-i\omega_m t} \sum_n k_n A_n^m e^{i\mathbf{k}_n \cdot \mathbf{r}} \cos^2 \phi_n\right], \quad (S22)$$

$$\varepsilon_{xy} = -\mathrm{Im}\left[e^{-i\omega_m t} \sum_n k_n A_n^m e^{i\mathbf{k}_n \cdot \mathbf{r}} \cos \phi_n \sin \phi_n\right]. \quad (S23)$$

Substituting the solution presented here into (S10), one obtains $L$ for a given eigenmode $\{A_0^m, A_1^m, A_2^m\}$ at position $\boldsymbol{r}$

$$L_m(\boldsymbol{r}) = \omega \sum_{n \neq n'} k_n k_{n'} \mathrm{Im}\left[\overline{A_n^m} A_{n'}^m e^{i(\mathbf{k}_{n'} - \mathbf{k}_n) \cdot \mathbf{r}}\right] \cos \phi_n \cos \phi_{n'} \sin \phi_{n'n}. \quad (S24)$$

Since $\mathbf{k}_n - \mathbf{k}_{n'}$ is a reciprocal vector, the average of $L$ over a real space unit cell always vanishes. The total $P_{\mathrm{chiral}}$ will therefore be zero for paramagnets with a spatially uniform distribution of spins. It indicates the significance of the periodically tuned magnon-phonon coupling in generating the sizable nonreciprocity observed in the experiment.

To capture the inhomogeneous magnon-phonon coupling in the unit cell with a minimal number of parameters, we consider honeycomb lattice formed by islands on A- and B-sites, with spatially homogenous strains and magnetization (like a macro-spin) within each island. The strain components are approximated by their values at the island center, neglecting the spatial variation within the islands. Then the shape of the islands is not taken into account and the remaining parameters are the areas of the islands. It is known that A- and B-sites usually have opposite senses of displacement rotation [11], i.e., one is clockwise while the other is counterclockwise. Therefore, if the contributions of A- and B-sites are summed with the same magnetic ground state, magnonic phase and amplitude, the nonreciprocity will be largely cancelled, in contrast with the large nonreciprocity observed in the experiment. In reality, the magnonic phase and amplitude at the K valley can be significantly different at A- and B-sites, and the magnetic ground state can be also different due to the different anisotropy at the two sites. These effects are responsible for the large nonreciprocity. As we are treating the magnon-phonon coupling locally (i.e., ignoring the magnetic coupling between different sites), and solving the magnonic eigenstates in the complex structure will be rather difficult, we take the latter effect into account by assuming that only A-site is near the resonance condition, and the B-site contribution is ignored for simplicity. We rewrite Eq. (S8) as

$$\Delta P = \frac{\alpha \gamma b^2 \omega^2}{M_S} \frac{\omega^2 \left(|\varepsilon_{xx} \sin 2\theta|^2 + 4|\varepsilon_{xy} \sin \theta|^2\right) - i\omega_F L \sin 2\theta \sin \theta}{\left(\omega_F^2 - \omega^2\right)^2 + \alpha^2 \omega_F^2 \omega^2}, \quad (S25)$$

where $\omega_F$ is the FMR frequency, under the approximation that the Ni magnetization with resonance frequency $\omega_F$ is modelled by the paramagnetic moment under the same resonance frequency and with the same tilting angle $\theta$.

For each eigenmode $\{A_0^m, A_1^m, A_2^m\}$, we calculate the strain polarization $L$ on A-site from Eq. (S24), and the absorption contribution from (S25). For Fig. 4 in the main text, we use $\theta = 54.7°$, $\omega_F/2\pi = 1.8$ GHz.

We then consider the excitation of the eigenmodes $\{A_0^m, A_1^m, A_2^m\}$ by the incident wave $\{1, 0, 0\}$. In the usual scattering problem, one assumes a constant flux of the incident wave, and try to compute the steady state pattern after an infinite number of scatterings at the boundaries have occurred. Note that in this process, the wavevector is not conserved. In our problem, this programme is not viable, not only because we do not know the model of the boundary conditions but also because the translation symmetry is broken along $y$ as well. Thus we take a completely different point of view. Consider only the patterned region and suppose there was a spatial profile of the unperturbed incident wave at an initial time. This initial profile is then assumed to evolve according to the time-domain version of (S18)-(S20). The general solution can be expanded in terms of (S21) with various $k_0$, $m$, but by orthogonality of plane waves, only the solution with $k_0$ equal to the incident wavevector will have nonzero initial amplitudes:

$$\boldsymbol{u}_{in} = \sum_{m=1}^{3} C_m \left( A_0^m e^{i\boldsymbol{k}_0 \cdot \boldsymbol{r}} \frac{\boldsymbol{k}_0}{k_0} + A_1^m e^{i\boldsymbol{k}_1 \cdot \boldsymbol{r}} \frac{\boldsymbol{k}_1}{k_1} + A_2^m e^{i\boldsymbol{k}_2 \cdot \boldsymbol{r}} \frac{\boldsymbol{k}_2}{k_2} \right). \text{(S26)}$$

Assuming the eigenvectors are normalized by

$$\left|A_0^m\right|^2 + \left|A_1^m\right|^2 + \left|A_2^m\right|^2 = 1, \text{(S27)}$$

and writing the initial profile as

$$\boldsymbol{u}_{in} = C e^{i\boldsymbol{k}_0 \cdot \boldsymbol{r}} \frac{\boldsymbol{k}_0}{k_0}, \text{(S28)}$$

one infers $C_m = C \overline{A_0^m}$. Note that modes with different $m$ will evolve at different frequencies. Therefore, in computing the time-averaged power absorbed by the magnetic resonance, one may discard the cross terms between different modes. Denoting the power by a normalized mode $m$ by $\Delta P_m$, one then obtains the total power absorbed starting from the incident wave with $k_0$ as

$$\Delta P_T = \left|C\right|^2 \sum_{m=1}^{3} \left|A_0^m\right|^2 \Delta P_m. \text{(S29)}$$

In the same band, at K and K' valley, $L$ has opposite values, leading to the different scattering based on (S29), i.e., the nonreciprocity due to the valley-selective phonon scattering,

$$\eta = \frac{P_T(\text{K}) - P_T(\text{K}')}{P_T(\text{K}) + P_T(\text{K}')}. \text{(S30)}$$

Since $\sin 2\theta \sin \theta$ changes sign when $\theta$ is shifted by $\pi$, there is $\Delta P_T(\theta + \pi) = -\Delta P_T(\theta)$, resulting in a reversed nonreciprocity when the magnetic field is reversed.

Next, to map the gap in the transmission spectra, we calculate the transmitting energy flow in the $x$ direction. Since we base our consideration on the spatial profile and not time scales and frequencies, calculations of transmission spectra without magnetic absorption cannot be made directly. Hence, we again consider the following get-around scenario. Suppose the conversion from the unperturbed to perturbed is efficient so that it occurs near the boundary and within a few spatial periods into the patterned area. Then the transmitted wave will be largely determined by how fast the energy is transmitted in the positive $x$ direction, i.e. $x$ component of the energy flux. Note that the slower the flow of energy is, the more energy is dissipated in the patterned area assuming the relaxation rate is the same. Applying the Noether's theorem to the elastic wave Lagrangian density

$$\mathcal{L} = \frac{1}{2}\left\{\rho\left|\frac{\partial \boldsymbol{u}}{\partial t}\right|^2 - (\lambda+\mu)(\nabla\cdot\boldsymbol{u})^2 + \mu\boldsymbol{u}\cdot\nabla^2\boldsymbol{u}\right\}, \quad (S31)$$

the energy flux $T = \langle \frac{\partial \mathcal{L}}{\partial(\partial_x u_i)}\frac{\partial u_i}{\partial t}\rangle$ is obtained as

$$T = -\langle(\lambda+2\mu)\partial_x u_x \partial_t u_x + \mu \partial_x u_y \partial_t u_y\rangle, \quad (S32)$$

where the angled bracket represents time-averaging. Replacing (S21) into (S32) and integrating in the unit cell, the $T$ of eigenmode $\{A_0^m, A_1^m, A_2^m\}$ is obtained as

$$T_m = 2\operatorname{Re}\left[\sum_n k_n |A_n^m|^2 \{(\lambda+2\mu)\cos^2\phi_n + 2\mu\sin^2\phi_n\}\cos\phi_n\right], \quad (S33)$$

Similar as (S29), the total $T$ of the mixed state excited by the incident wave is estimated by

$$T = |C|^2 \sum_{m=1}^{3} |A_0^m|^2 T_m. \quad (S34)$$

6.2 M-point in Zigzag-like lattice

We now turn to the Zigzag-like lattices which do break inversion symmetry but lack valley degrees of freedom. The coordinates of site A and B are taken to be $(0,\sqrt{2}a/6,0)$ and $(0,-\sqrt{2}a/6,0)$, with scattering amplitudes $s_1$, $s_2$, respectively [Fig. S7(a)]. The basic translation vectors are

$$\boldsymbol{a}_1 = \frac{a}{\sqrt{2}}\begin{pmatrix}1\\-1\\0\end{pmatrix}, \boldsymbol{a}_2 = \frac{a}{\sqrt{2}}\begin{pmatrix}1\\1\\0\end{pmatrix}, \quad (S35)$$

for which the reciprocal lattice vectors are

$$\boldsymbol{b}_1 = \frac{2\pi}{\sqrt{2}a}\begin{pmatrix}1\\-1\\0\end{pmatrix}, \boldsymbol{b}_2 = \frac{2\pi}{\sqrt{2}a}\begin{pmatrix}1\\1\\0\end{pmatrix}. \quad (S36)$$

Let $\boldsymbol{k}_M = (\boldsymbol{b}_1 + \boldsymbol{b}_2)/2$, which yields $\Omega = \{-\boldsymbol{b}_1, -\boldsymbol{b}_2, -\boldsymbol{b}_1-\boldsymbol{b}_2\}$, as shown in Fig. S7(b). $k_M$ is defined as the length of the wavevector at M. We label the wavevectors $\boldsymbol{k}_0 = \boldsymbol{k}_{K_1} + \boldsymbol{q}, \boldsymbol{k}_1 = \boldsymbol{k}_0 - \boldsymbol{b}_1, \boldsymbol{k}_2 = \boldsymbol{k}_0 - \boldsymbol{b}_2, \boldsymbol{k}_3 = \boldsymbol{k}_0 - \boldsymbol{b}_1 - \boldsymbol{b}_2, |\boldsymbol{q}| \ll 2\pi/a$ and obtain

$$\omega^2 A_0 = c_P k_0^2 A_0 + \rho_a^{-1}(\lambda_{\boldsymbol{b}_1} + 2\mu_{\boldsymbol{b}_1}\cos^2\phi_{01})k_0 k_1 A_1 + \rho_a^{-1}(\lambda_{\boldsymbol{b}_2} + 2\mu_{\boldsymbol{b}_2}\cos^2\phi_{02})k_0 k_2 A_2$$
$$+\rho_a^{-1}(\lambda_{\boldsymbol{b}_1+\boldsymbol{b}_2} + 2\mu_{\boldsymbol{b}_1+\boldsymbol{b}_2}\cos^2\phi_{03})k_0 k_3 A_3, \quad (S37)$$

$$\omega^2 A_1 = c_P k_1^2 A_1 + \rho_a^{-1}(\lambda_{-\boldsymbol{b}_1} + 2\mu_{-\boldsymbol{b}_1}\cos^2\phi_{01})k_0 k_1 A_0 + \rho_a^{-1}(\lambda_{\boldsymbol{b}_2-\boldsymbol{b}_1} + 2\mu_{\boldsymbol{b}_2-\boldsymbol{b}_1}\cos^2\phi_{12})k_1 k_2 A_2$$
$$+\rho_a^{-1}(\lambda_{\boldsymbol{b}_2} + 2\mu_{\boldsymbol{b}_2}\cos^2\phi_{13})k_1 k_3 A_3, \quad (S38)$$

$$\omega^2 A_2 = c_P k_2^2 A_2 + \rho_a^{-1}(\lambda_{-\boldsymbol{b}_2} + 2\mu_{-\boldsymbol{b}_2}\cos^2\phi_{01})k_0 k_1 A_0 + \rho_a^{-1}(\lambda_{\boldsymbol{b}_1-\boldsymbol{b}_2} + 2\mu_{\boldsymbol{b}_1-\boldsymbol{b}_2}\cos^2\phi_{12})k_1 k_2 A_1$$
$$+\rho_a^{-1}(\lambda_{\boldsymbol{b}_1} + 2\mu_{\boldsymbol{b}_1}\cos^2\phi_{23})k_2 k_3 A_3, \quad (S39)$$

$$\omega^2 A_3 = c_P k_3^2 A_3 + \rho_a^{-1}(\lambda_{-\boldsymbol{b}_1-\boldsymbol{b}_2} + 2\mu_{-\boldsymbol{b}_1-\boldsymbol{b}_2}\cos^2\phi_{03})k_0 k_3 A_0 + \rho_a^{-1}(\lambda_{-\boldsymbol{b}_2} + 2\mu_{-\boldsymbol{b}_2}\cos^2\phi_{12})k_1 k_3 A_1$$
$$+\rho_a^{-1}(\lambda_{-\boldsymbol{b}_1} + 2\mu_{-\boldsymbol{b}_1}\cos^2\phi_{23})k_2 k_3 A_2, \quad (S40)$$

The notations are analogous to the previous subsection. We note $\cos\phi_{01} \approx \cos\phi_{02} \approx \cos\phi_{13} \approx \cos\phi_{23} \approx 0$ and $\cos\phi_{03} \approx \cos\phi_{12} \approx -1$. The solution is of the form

$$\boldsymbol{u} = \mathrm{Re}\left[ e^{-i\omega_m t}\left( A_0^m e^{i\boldsymbol{k}_0 \cdot \boldsymbol{r}} \frac{\boldsymbol{k}_0}{k_0} + A_1^m e^{i\boldsymbol{k}_1 \cdot \boldsymbol{r}} \frac{\boldsymbol{k}_1}{k_1} + A_2^m e^{i\boldsymbol{k}_2 \cdot \boldsymbol{r}} \frac{\boldsymbol{k}_2}{k_2} + A_3^m e^{i\boldsymbol{k}_3 \cdot \boldsymbol{r}} \frac{\boldsymbol{k}_3}{k_3} \right) \right]. \quad \text{(S41)}$$

and the strain tensor components can be computed similarly to the previous subsection. It is easy to get $f_{b_1} = s_1 t_1 + s_2 t_2$, $f_{b_2} = s_1 t_2 + s_2 t_1$, $f_{b_1 - b_2} = s_1 t_3 + s_2 t_4$, $f_{b_1 + b_2} = s_1 + s_2$, where $t_1 = \exp(\pi i/3)$, $t_2 = \exp(-\pi i/3)$, $t_3 = \exp(2\pi i/3)$, $t_4 = \exp(-2\pi i/3)$. The values of $s_1$ and $s_2$ are kept the same as the honeycomb lattice. Following the same procedure as in the honeycomb lattice, one can obtain the nonreciprocity in the Zigzag-like lattice. The result is shown in Fig. S7(c). Due to the translation symmetry identifying $+k$ and $-k$ at M point, the nonreciprocity vanishes for $k = k_M$. When $k$ moves away from $k_M$, there exists no relevant symmetry, and a nonreciprocity of the order of 1% emerges. These calculations based on the plane-wave scattering model qualitatively reproduce the experimental results.

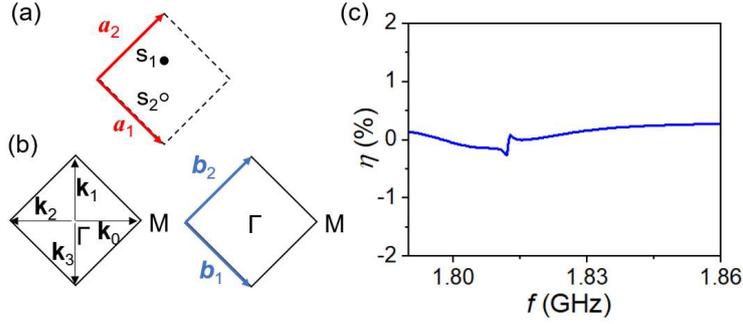

FIG. S7. (a) Schematic of the real space scattering model constructed by delta-scattering function for the Zigzag-like lattice. (b) Brillion zone and the relevant reciprocal vectors in the model. (c) The calculated nonreciprocity as a function of the incident $k$ vector.

The parameters used for the elastic wave scattering model are summarized as follows.

| Parameter | value (expression) |
| --- | --- |
| $k_K$, $k_M$ ($2\pi$ μm$^{-1}$) | 0.5 |
| $c_P$ (m/s) | 3630 |
| $c_S$ (m/s) | 1815 |
| $s_1$ | −0.01 |
| $s_2$ | −0.003 |
| $\alpha$ | 0.05 |
| $\rho_a^{-1}\lambda_a$ | $c_P^2 - 2c_S^2$ |
| $\rho_a^{-1}\mu_a$ | $c_S^2$ |
| $f_G$ | obtained from $s_1$, $s_2$ and lattice structure |
| $\lambda_G$ | $\lambda_a f_G$ |
| $\mu_G$ | $\mu_a f_G$ |
| $\theta$ | 54.7° |
| $\omega_F/2\pi$ | 1.8 GHz |

TABLE S1. Parameters for calculation in the plane wave scattering model.